\documentclass[12pt,preprint,flushrt]{aastex}
\doublespace
\begin{document}

\title{Discovery of X-ray QPOs from an Ultra-luminous X-ray Source in M82: 
Evidence Against Beaming}
\author{Tod E. Strohmayer and Richard F. Mushotzky}
\affil{Laboratory for High Energy Astrophysics, \\ NASA's Goddard Space Flight 
Center, Greenbelt, MD 20771}
\email{stroh@clarence.gsfc.nasa.gov}
%\authoraddr{Laboratory for High Energy Astrophysics, Mail Code 662, NASA/GSFC
%Greenbelt, MD 20771}

\begin{abstract}

We report the discovery with the EPIC CCD cameras onboard XMM-Newton of a 54 
mHz quasiperiodic oscillation (QPO) in the $> 2$ keV X-ray flux from an
ultraluminous X-ray source (ULX) in the starburst galaxy M82. This is the 
first detection of a QPO in the X-ray flux from an extra-Galactic
ULX, and confirms that the source is a compact object. Based on the 
QPO strength and previous Chandra observations it appears likely that the QPO 
is associated with the most luminous object in the central region of M82, 
CXOM82 J095550.2+694047, however, XMM imaging alone is not sufficient to 
unambiguously confirm this. The other plausible candidate 
is CXOM82 J095551.1+694045, however, the QPO luminosity is comparable to the 
{\it peak} luminosity of this object in Chandra observations, which argues 
against it being the source of the QPO. The QPO had a centroid frequency of 
$54.3 \pm 0.9$ mHz, a coherence $Q \equiv \nu_0 / \Delta\nu_{fwhm} \approx 5$, 
and an amplitude (rms) in the 2 - 10 keV band of $8.5 \%$. Below 0.2 Hz the 
power spectrum can be described by a power-law with index $\approx 1$, and 
amplitude (rms) of $13.5 \%$. The X-ray spectrum requires a curving continuum, 
with a disk-blackbody (diskbb) at $T = 3.1$ keV providing an acceptable fit. 
A broad Fe line centered at 6.55 keV is required in all fits, but the 
equivalent width (EW) is sensitive to the continuum model. There is no 
evidence of a reflection component. The implied bolometric luminosity is 
$\approx 4 - 5 \times 10^{40}$ ergs s$^{-1}$. Archival Rossi X-ray 
Timing Explorer (RXTE) pointings at M82 also show evidence for QPOs in the 
50 - 100 mHz frequency range.  We discuss the implications of our findings for 
models of ULXs. 

\end{abstract}


\vfill\eject

\section{Introduction}

Many nearby galaxies harbor highly luminous point-like X-ray sources which 
are not associated with the nuclear region of the galaxy. 
Their existence has been known for some time and was first 
inferred from observations with the EINSTEIN observatory (Fabbiano 1988). 
These ultra-luminous X-ray sources (ULXs) can have apparent isotropic X-ray
luminosities of $10 - 1000 \times$ the Eddington limit for a canonical neutron 
star.  Although some of them may be highly luminous young 
supernova remnants, the presence of significant time variability on timescales
of days to years argues for a compact accretor (see Ptak \& Griffiths 1999; 
Fabbiano \& Trinchieri 1987; Schlegel 1994). These findings have led to the 
suggestion that they may represent a new class of intermediate 
mass ($50 - 1000$ $M_{\odot}$) black holes (IMBHs, Colbert \& Mushotzky 1999; 
Makishima et al. 2000), perhaps formed as a result of binary interactions in 
dense stellar environments (see for example, Portegies Zwart \& McMillan
2002). 

Current models for these objects  
center around two alternatives; (1) They are massive (50 - 1000 
$M_{\odot}$) black holes radiating at or near the Eddington limit, or 
(2) they are more or less ``normal'' (ie. stellar mass) accreting compact 
binaries whose energy loss appears super-Eddington because it is beamed. 
Each of these explanations is not without its difficulties. 
For example, if they are massive black holes, then standard accretion disk 
theory suggests that they should have lower disk temperatures than stellar mass
black holes. Spectroscopy with ASCA (Makishima et al. 2000), however, does 
not support this conclusion unless, perhaps, they are rapidly 
rotating Kerr holes (see, however, a recent report by Miller et al. 2002 
suggesting a cooler disk in one source). Moreover, formation scenarios 
for IMBHs are challenging, as is the need to feed the black hole at the 
requisite accretion rates. Because of some of these difficulties, 
King et al. (2001) have argued for an association with ``standard'' X-ray 
binaries whose emission is beamed geometrically by factors of $\sim 10 - 100$ 
(see also Zezas \& Fabbiano 2002). 
Recently, Grimm, Gilfanov \& Sunyaev (2002) have demonstrated a connection 
between high mass X-ray binaries (HMXBs) and star formation. This association 
supports the putative link between star formation and ULXs, and suggests they 
may reflect the high mass end of the luminosity function of HMXBs. However, 
optical searches for the high mass counterparts of the ULXs have not so far
been very productive, and details of how and indeed if sufficient beaming 
occurs at high mass accretion rates are not well understood theoretically 
(Madau 1988; Kubota, Done \& Makishima 2002).  

There is now an extensive amount of information on the timing properties 
of Galactic black holes candidates (BHCs). In particular, RXTE observations 
have discovered QPOs with frequencies ranging from 
0.001 to 450 Hz (see Remillard et al. 2002 for a brief summary). Many of 
these QPOs are strongly correlated with spectral states. 
If ULXs are extragalactic analogs of the stellar mass compact binaries in our 
own galaxy, then they should show some of the same timing properties and 
spectral correlations. Although these objects are faint by Galactic BHC 
standards, as we demonstrate here, the brightest sources are good candidates 
for low frequency ($\sim 1 - 100$ mHz) QPO searches with large area, imaging 
instruments, such as the EPIC CCDs onboard XMM/Newton. This possibility has 
led us to examine the timing properties of some of the brighter ULXs which 
have been observed with XMM/Newton and Chandra.

One of the brightest ULXs is the source in M82 (CXOM82 J095550.2+694047, source
7 in Matsumoto et al. 2001, hereafter M82 X-1). Based on ASCA data Ptak \& 
Griffiths (1999) found both spectral and temporal variability in the hard 
X-ray flux from M82, and suggested that the hard X-ray flux is from a compact 
object of $\approx 500 M_{\odot}$. Recent Chandra HRC observations confirm 
that the ULX is not associated with the dynamical center of M82 nor is it 
associated with any radio AGN candidate or optically bright counterpart 
(Kaaret et al. 2001; Matsumoto et al. 2001). 
Here we report timing and spectroscopy of this object utilizing XMM/Newton and 
RXTE archival data. With XMM we detect a 54 mHz QPO in the hard X-ray flux 
from this object. We also find indications of 50 - 100 mHz QPOs from RXTE 
observations. We present evidence for a broad Fe K$\alpha$ emission feature 
in the XMM spectrum. We discuss the implications of these findings for the 
nature of ULXs.

\section{XMM Observations and Data Analysis}

XMM/Newton observed M82 for 30 ksec on May 5, 2001 at 09:19:40 (UT). These
data are now in the public XMM archive.  For our study we used only the EPIC
data. We used the standard SAS version 5.3.3 tools to filter and extract 
images and event tables for both the PN and MOS cameras.  These instruments
were in PrimeFullWindow mode with the medium blocking filter. We began by 
extracting images from both the PN and MOS cameras.  The central region of
M82 has been resolved by Chandra (see Kaaret et al. 2001; Matsumoto et al. 
2001). The PN and MOS images are dominated by a bright point source whose
location is consistent with M82 X-1. At XMM's spatial resolution, however, the 
field is crowded by several fainter sources resolved by Chandra. We have 
reexamined the available Chandra ACIS and HRC imaging and we cannot rule out 
some contribution to the XMM flux from several nearby sources (sources 4, 5 and
6 in Matsumoto et al. 2001). Of these, sources 4 and 6 were not seen to vary 
with Chandra and were always {\it at least} a factor of $\approx 10$ fainter 
than M82 X-1. Source 5 was found to be variable with Chandra and at its 
brightest was a factor 3.4 fainter than M82 X-1 at its faintest.  

Previous measurements indicate that the central region of M82 has substantial 
diffuse, soft thermal emission. We see this diffuse component in both the 
EPIC images and spectra (see Figure 4). Inside an 18'' extraction radius the 
emission from the thermal component is $< 10 \%$ of the point source at 
$E >$ 2 keV, therefore, to increase the sensitivity to variability from the 
compact source we use only the hard X-rays ($> 2$ keV) for our timing study. 
To investigate the source's temporal variability we extracted events from the
PN and MOS cameras within a 18'' circular region around the bright point
source.  To make the most sensitive search we combined data from the PN and 
MOS cameras into a single lightcurve.  Figure 1 shows the 2 - 10 keV lightcurve
from the PN and MOS. The average 2 - 10 keV countrate in the PN + MOS is 
3.18 s$^{-1}$.

\subsection{Power Spectral Timing Analysis}

We calculated a single FFT power spectrum using a lightcurve sampled at 1/2 s,
yielding a 1 Hz Nyquist frequency. We used only the time interval for which
both the PN and MOS were operating. This gave a continuous exposure of 
$\approx 27$ ksec. Figure 2 shows the 2 - 10 keV power spectrum rebinned by a 
factor of 128 in frequency space (frequency resolution 4.7 mHz) for the 
PN + MOS (lower curve), the PN only (middle curve), and the MOS only (sum of 
MOS1 and MOS2, top curve). There is a prominent QPO peak centered near 54 mHz 
in all three power spectra. To assess the significance of the QPO we fit the 
power spectrum using a model composed of a constant plus a power law and a 
Lorentzian. This model provides a good fit, giving a minimum $\chi^2 = 185$ 
with 206 degrees of freedom (dof), and is shown in Figure 2 as the solid curve 
through the PN + MOS spectrum. If the Lorentzian (QPO) component is excluded 
from the model, the fit is unacceptable, with a minimum $\chi^2 = 255$. This 
gives a $\Delta\chi^2 = 70$ for the additional 3 parameters of the QPO 
component. The significance of the additional parameters can be estimated 
with the F-test, and gives a probability of $\approx 3 \times 10^{-14}$, which 
strongly indicates the need for the QPO component. 

To make another estimate of the significance of the QPO we used the power law
parameters from our best fit model to rescale the power spectrum so that the 
local mean power in the vicinity of the QPO was 2, the level expected for 
pure counting noise in the spectrum. The probability of obtaining by chance 
the highest power, $P_{max} = 3.28$, in the QPO profile is then just the 
probability of obtaining a power $P = P_{max} \times 128$ from the $\chi^2$ 
distribution with $2\times 128 = 256$ degrees of freedom (this because we 
averaged 128 independent powers in rebinning the spectrum). This gives a single
trial probability of $4.5 \times 10^{-10}$ for the highest power seen in the 
QPO profile. Multiplying by the number of powers searched, $N_{trials} = 212$, 
in the power spectrum then gives a significance of $9.5 \times 10^{-8}$, which 
is about a $6\sigma$ detection.  Based on this and the F-test estimate above, 
we are very confident in the QPO detection.

When we construct lightcurves using the PN events at higher 
sampling frequencies and compute power spectra we clearly see the 13.63 Hz 
(73 ms) frame sampling rate, its harmonics, and aliases of some of the 
harmonics. The 54 mHz QPO cannot be due to an alias of this sampling frequency.
The aliases of its first several harmonics are not at or near the measured QPO 
frequency, and any aliased power from higher harmonics is greatly suppressed 
by the time binning. These features are also narrow, 
whereas the QPO is not. Similar arguments hold for the MOS analysis. 
When we analyse only photons with energies less than 2 keV the QPO is not 
detected, indeed, there is no indication of any significant variability. 
This is consistent with the conclusion that most of the soft X-ray flux is
from the diffuse emission in the XMM beam, and also provides additional 
evidence that the QPO is not due to some instrumental signature. 

Our best power spectral model includes a power law component, 
$A \nu^{-\alpha}$, with $\alpha = 1 \pm 0.12$, and an integrated 
variability (rms) of $13.5 \%$ from 0.1 mHz to 1 Hz. The QPO has a centroid 
frequency of $54.4 \pm 0.9 $ mHz,a width $\Delta\nu_{fwhm} = 11.4 \pm 2$ mHz, 
and an amplitude (rms) of $8.4 \%$ in the 2 - 10 keV band. We searched for 
energy and time dependence of the QPO amplitude, but did not find any strong 
dependence. However, due to the relatively low signal to noise ratio,
the data are not particularly sensitive to either of these effects.

\subsection{RXTE Timing Analysis}

Based on our XMM detection we decided to search archival RXTE observations 
for similar timing features. Indeed, $\sim 100$ ksec of monitoring observations
were carried out on M82 with RXTE from February to November of 1997 (Proposal 
ID 20203; see Gruber \& Rephaeli 2002). These were typically 3 - 4 ksec 
pointings utilizing 3 of the 5 Proportional Counter Array (PCA) detectors.  
We extracted lightcurves with a 128 Hz sampling rate using only the top xenon 
layer and events in the 2 - 20 keV energy band. Here we only summarize results 
from several of the observations with indications of QPOs in the 50 - 100 mHz 
range. A complete analysis of the timing properties of all the observations 
will be presented in a sequel. Figure 3 shows power spectra from three 
observations with QPO detections. Each power spectrum is labelled with the 
corresponding observation ID. Model fits including a power law and a Lorentzian
are also shown.  

The two most significant QPOs are in the 20203-02-04-00 and 20203-02-02-00 
observations. These QPOs have frequencies of $107 \pm 3$ mHz, and $51 \pm 2$ 
mHz, and F-test significances of $1.7 \times 10^{-6}$ and $3.2 \times 10^{-6}$,
respectively. To determine the QPO amplitudes we first computed background
countrates using the RXTE/PCA background models. This gave 2 - 20 keV source 
counting rates of $\approx 8 - 9$ s$^{-1}$, and corresponding amplitudes (rms) 
of about 8 \% for each QPO.  The QPO frequencies, amplitudes and coherences
inferred from the RXTE data are very similar to those derived from the XMM
data. As we discuss below, the flux levels inferred for the RXTE observations 
with QPOs are consistent with the XMM flux level. The similar flux levels and 
QPO properties suggest that the source was in a similar state (see Gruber \& 
Rephaeli 2002). However, the RXTE monitoring also found the 
source at a flux level higher by a factor of 3 on 4 occasions. 
Interestingly, these observations did not reveal any QPOs. 
Either the QPO source properties changed in a way suggestive of a 
correlation between spectral states and timing properties or perhaps an 
unrelated source or sources brightened and decreased the QPO sensitivity. 
With the RXTE data alone it is not possible to decide between these two 
alternatives.

\subsection{Energy Spectral Analysis}

We proceed under the assumption that a single source (most likely M82 X-1) 
dominates the XMM flux, but we comment below on the possible effects of source 
confusion. We obtained the spectrum by extracting a region with radius of 
18". This extraction region has
extensive diffuse emission which is spatially resolved by Chandra (see for 
example, Kaaret et al. 2001). Fitting the Chandra spectrum of the bright 
point source one derives an effective column density of $\ 0.5 - 0.9 \times 
10^{21}$ cm$^{-2}$ depending on the continuum model used. We then fit the 
XMM data in the 3 - 10 keV band from all 3 instruments. We find that a power 
law model is considerably worse than a disk blackbody (diskbb) or comptonized 
(compst) model (minimum $\chi^2 = 2907$ vs 2385 or 2246, with 2021 dof), with 
the column density fixed to the range allowed by the Chandra data. Further, 
no significant reduction in $\chi^2$ is obtained for more complex continuum
models (e.g. the addition of a power law to a diskbb model) often used to 
model Galactic BHCs. However, there are broad residuals in the spectrum no
matter what continuum model is used. These residuals can be modelled by a 
variety of line shapes, including a relativistic broad line 
(the Laor model in XSPEC), the standard diskline model with 
extreme parameters ($q = -5.8$ with the diskbb continuum), or a gaussian. 
These fits indicate that much of the flux in the line comes from regions close 
to the black hole. The derived parameters for the line are, however, sensitive 
to the continuum model. The most ``curved'' continuum, the compst model, 
has the lowest EW (230 eV) while the powerlaw model (which fits the continuum 
poorly) has an EW of 1300 eV. There is no evidence for a reflection component.
Fitting a simple power law model in the 5 - 10 keV band (and not including 
any line emission) overpredicts the flux in the 3 - 5 keV band by a factor of 
$\approx 1.6$ and does not account for the obvious line. The line is centered 
at 6.55 keV ($1\sigma$ range of 6.48 - 6.60 keV) and has a gaussian width of 
0.33 (0.26 - 0.43) keV. Figure 4 shows the total spectrum of M82 inside the 
18'' extraction radius.

The derived diskbb temperature, $T_{in}$, is somewhat sensitive to the choice 
of column density, however, values less than 3 keV are strongly excluded. 
Chandra ACIS spectra of M82 X-1 which suffer no source 
confusion indicate $T_{in} > 2.25$, however the source was significantly 
fainter in these observations. There is a known positive correlation 
of $T_{in}$ with luminosity for ULXs (Makishima et al. 2000), so that the
higher temperature inferred from the XMM data is not inconsistent with
the notion that the XMM spectrum is dominated by M82 X-1. 
The 2 - 10 keV flux is $2.1 \times 10^{-11}$ ergs cm$^{-2}$ s$^{-1}$. The 
bolometric flux of the compst model is $3.3 \times 10^{-11}$ ergs 
cm$^{-2}$ s$^{-1}$, while that of the diskbb model is $3.5 \times 10^{-11}$ 
ergs cm$^{-2}$ s$^{-1}$. Using a distance of 3.5 Mpc this gives bolometric 
luminosities of $\approx 4 - 5 \times 10^{40}$ ergs s$^{-1}$. Assuming 
a similar spectrum, the bolometric luminosity of the object 
during the 4 brightest RXTE observations is $3\times$ larger or 
$\sim 1 \times 10^{41}$ergs s$^{-1}$. 

\section{Discussion and Summary}

What is the source of the QPO and Fe line? The effective temperature of the 
diffuse emission near M82 X-1 from the Chandra data is too low to produce an
Fe K line. Moreover, it is unlikely that a diffuse process would produce a 
broad line. Unfortunately, the Chandra data for the point sources are not 
sensitive enough to detect the Fe K line seen in the XMM spectra. A number of 
arguments support the idea that M82 X-1 is the source of the QPO. 1) The XMM 
flux is consistent with the highest flux levels seen from M82 X-1 with Chandra.
2) The peak luminosity from Chandra observations of the other plausible 
candidate (source 5 in Matsumoto et al. 2001) was $\approx 3.5 \times 10^{39}$ 
ergs cm$^{-2}$ s$^{-1}$. This is comparable to the {\it luminosity} of the QPO,
$L_{qpo} \approx 0.085 \times 4 \times 10^{40} = 3.4 \times 10^{39}$ ergs 
cm$^{-2}$ s$^{-1}$. So, either source 5 increased dramatically in brightness 
above what was seen in previous observations, or it would need to be modulated 
at $\approx 100 \%$ of its peak brightness to explain the QPO. These 
alternatives seem unlikely to us. A simpler interpretation is that both the QPO
and Fe line are produced in the same object, M82 X-1, however, it will likely 
require simultaneous Chandra and XMM observations to resolve this issue 
definitively. 

The QPO discovery establishes beyond doubt the compact
nature of the source.  A firm upper limit to the size of the hard X-ray 
emission region is, $r_{source} < c/\nu_{QPO} = 2.8 \times 10^6$ km $\approx 
4 \ R_{\odot}$. If the highest QPO frequency is associated with 
the Kepler frequency at the innermost circular orbit around a Schwarzschild 
black hole, then the mass must be $M_{bh} < 1.87 \times 10^{4} M_{\odot}$. 
This establishes the ULX in M82 as a stellar-sized object and not an AGN. 

Several Galactic microquasars, for example, GRS 1915+105, 
GRO J1655-40 and XTE J1550-564, show low frequency QPOs with similar 
frequency and strength as the QPOs from X41.4+60 described here 
(see Morgan, Remillard \& Greiner 1997; Remillard et al. 1999; and Cui et al. 
1999). For example, GRS 1915+105 shows $\sim 10 - 100$ mHz QPOs at times when 
the source is in a ``bright'' state characterized by relatively modest broad 
band variability ($10 - 15 \%$ rms, see Morgan, Remillard \& Greiner 1997).
To the extent that they can be compared, the timing properties of GRS 1915+105 
in this state are similar to the properties of the ULX reported here.
The QPO frequencies, amplitudes and coherences are similar, as is the power 
law index and strength of the broad band variability.  The energy spectrum of 
GRS 1915+105 in such a state is also qualitatively similar to the inferred 
XMM spectrum, being fit by a relatively hot disk-blackbody component, 
although the inferred inner disk temperature is higher for M82 X-1.  
GRO J1655-40 has also shown $80 - 100$ mHz QPOs when the spectrum has 
high inferred disk temperatures (see Remillard et al. 1999; Sobczak et al. 
1999). 

These comparisons would seem to suggest that the M82 ULX may be an analog of 
the Galactic microquasars, and thus that its mass is not extreme (see 
Greiner, Cuby \& McCaughrean 2001), however, there are 
difficulties with this conclusion. Detection of strong, narrow QPOs provides 
evidence for the presence of a geometrically thin accretion flow 
(ie. a disk, see van der Klis 1995; Di Matteo \& Psaltis 1999), which 
argues against substantial beaming. For example, Madau (1988) has computed 
spectra from thick accretion disks which can have beaming factors approaching 
25. The radiation field along the rotation axis is greatly enhanced by the 
multiple scattering of photons off the walls of the inner, very hot and 
luminous, funnel-shaped region. The funnel dominates the emission from tori 
which are viewed close to the symmetry axis. The presence of a narrow QPO 
challenges such a scenario for the M82 source, since multiple 
scatterings would degrade and broaden any QPO variability. Warped thin disks 
can also produce collimation (see for example, Pringle 1997; Maloney, 
Begelman \& Pringle 1996), and is indeed predicted for objects near the 
Eddington limit, however, the amount of collimation/beaming, is relatively 
modest and is probably not sufficient to account for the factor of 50 - 100 
required for M82 X-1. We note that the $\approx 7$ hr periodic 
dipping/eclipsing modulation seen in a putative ULX in the Circinus Galaxy 
(Bauer et al. 2001) also argues against substantial beaming in this source. 

Galactic black hole binaries also show $\sim 0.8 - 3$ Hz QPO in the
``very-high'' state (VHS, van der Klis 1995; Cui et al. 1999; Morgan, 
Remillard \& Greiner 1997). These ``disk - corona'' oscillations are 
associated with a non-thermal (power law) spectral component (see Swank 2001 
for a review). These states show a characteristic flat-topped, broad band 
noise which breaks to a steeper power law at about the QPO frequency (see 
Morgan, Remillard \& Greiner 1997 for examples from GRS 1915+105). Could the 
QPOs in the M82 ULX be the analog of these $\sim 1$ Hz QPOs? Assuming that 
frequencies scale roughly as $1/M_{bh}$ suggests a mass of order $100 - 300 
M_{\odot}$ to obtain a 54 - 100 mHz QPO. However, the broad band variability 
below the QPO in X41.4+60 is really not flat-topped, and the lack of a 
non-thermal (power law) component also distinguishes the ULX from Galactic 
BHCs in this state. 

We are left with something of a conundrum; the presence of strong, narrow 
QPOs suggests that most of the flux in the 2 - 9 keV band originates 
from a disk, and thus the emission is not strongly beamed. However no present 
disk models can produce a color temperature as high as seen in M82 X-1. The 
inferred inner disk radius using the diskbb model is $\approx 35$ km (assuming 
$T_{in} = 3.1$ keV, a distance of 3.5 Mpc, an inclination angle of 
$45^{\circ}$, and no spectral hardening corrections). A Schwarzschild black 
hole of $4 M_{\odot}$ would have a last stable orbit this size, and further 
highlights the luminosity problem. This suggests that the diskbb model is 
probably unphysical, indeed, a comptonization model seems more plausible. 
Since it now appears that the radiation is not strongly beamed it is also 
difficult to explain the observed long term bolometric luminosity with an 
object of $< 300 \ M_{\odot}$ despite the apparent difficulties with such a 
scenario. The QPO frequency and the broad Fe K line are reminiscent of that 
seen in the Galactic microquasars in the VHS, but the continuum is rather 
different. We are thus left with the possibility that these objects show 
behavior not seen in any other black hole candidates, either in the Milkyway 
or as AGN.

Our results indicate that a combination of sensitive X-ray timing and 
spectroscopy can provide new insights into the nature of ULXs. 
Future missions with even larger collecting area, good spatial and
spectral resolution and timing capabilities (such as Constellation-X and XEUS),
should be able to open up a new window on the properties of black holes in 
the local universe.    

\acknowledgements

We thank Keith Arnaud, Craig Markwardt and Jean Swank for many helpful 
discussions and comments on the manuscript. 

\vfill\eject

\vfill\eject

\section{Figure Captions}

\figcaption[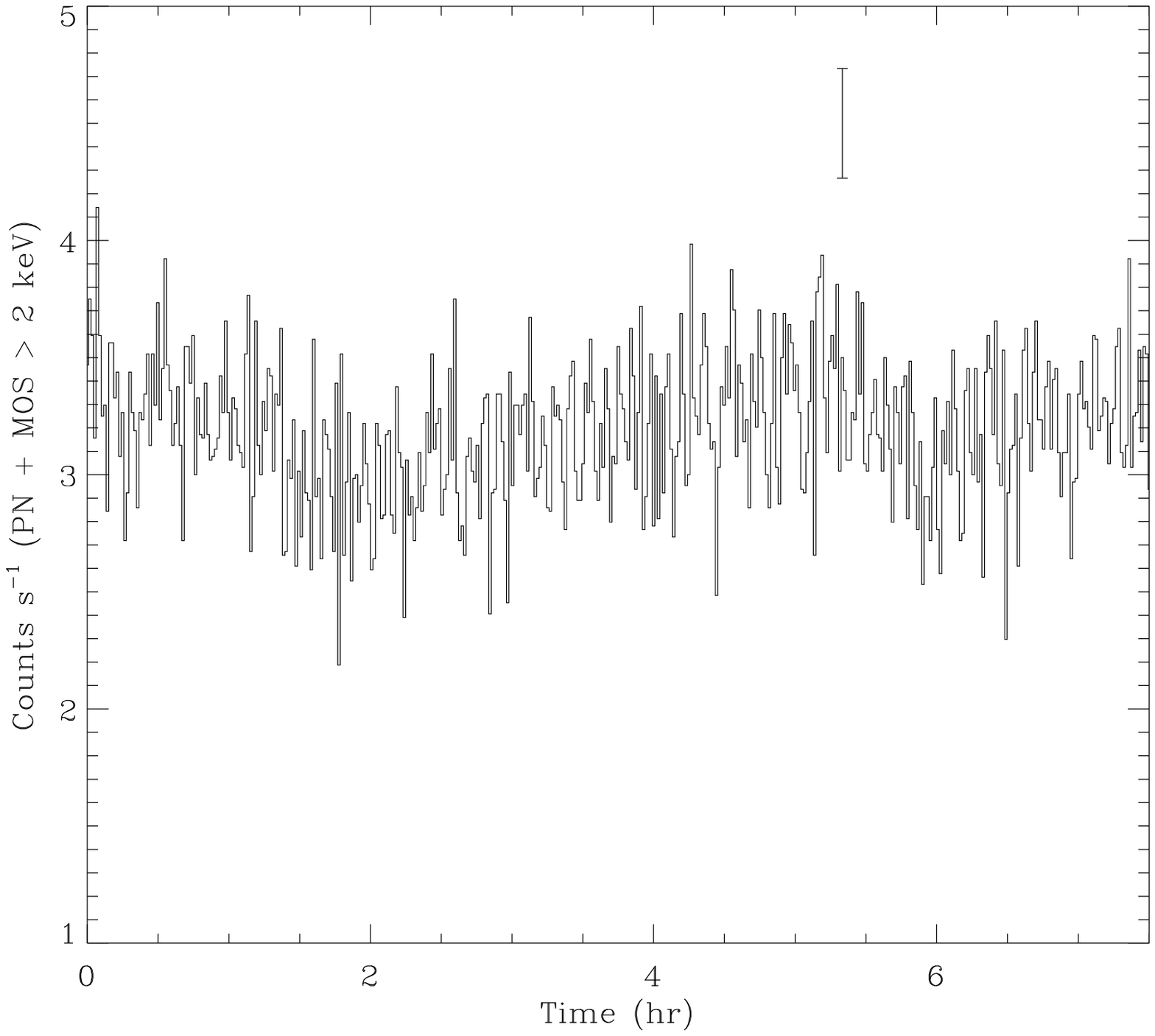]{EPIC PN + MOS lightcurve of the ULX near the central 
region of M82. The time bins are 64 seconds.  A characteristic error bar is 
also shown.  
\label{fig1}} 

\vskip 10pt

\figcaption[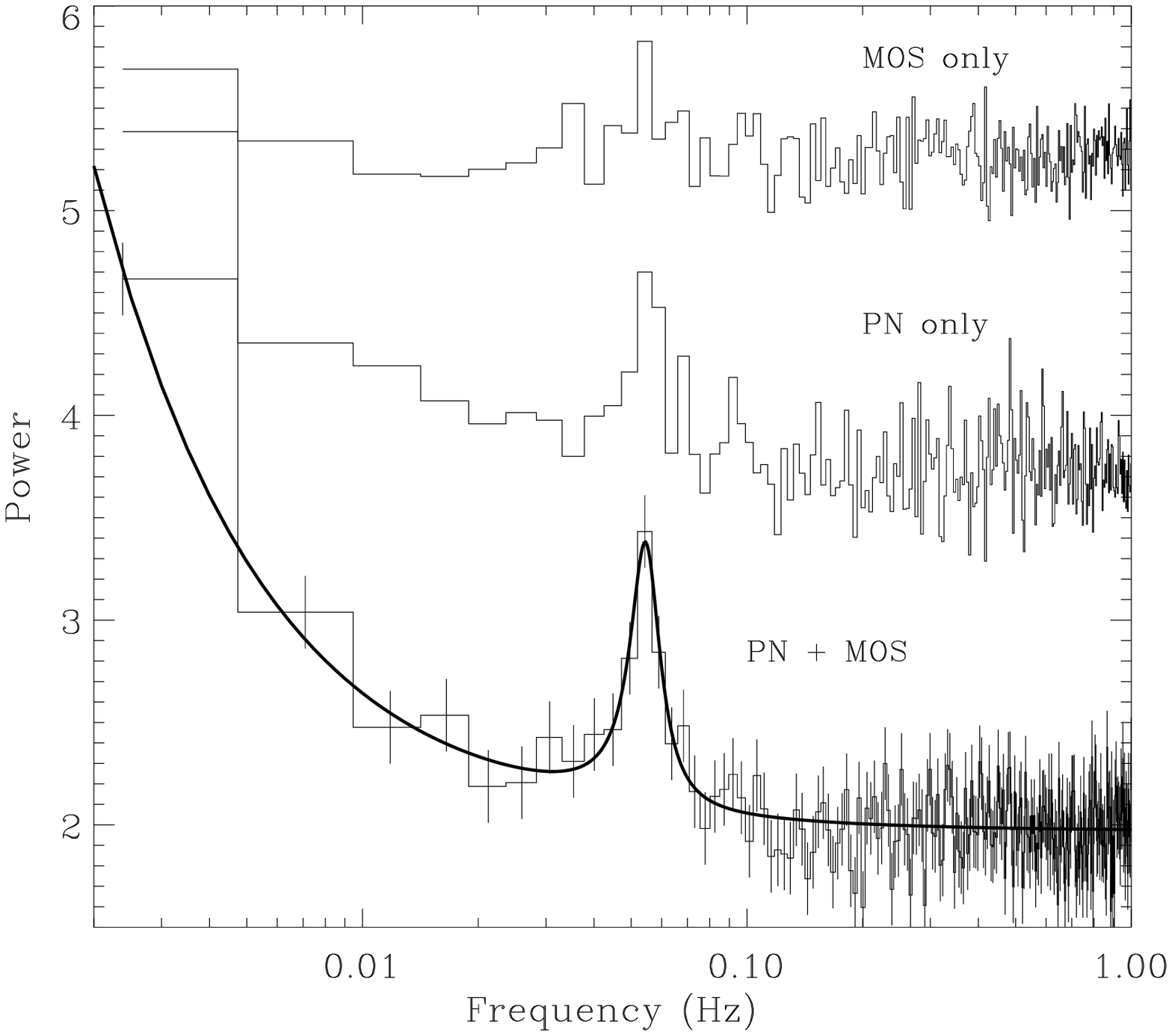]{Power spectrum of the EPIC $> 2$ keV data from 
X41.4+60. The Nyquist frequency is 1 Hz. The frequency resolution is 4.7 mHz. 
The Poisson level has not been subtracted. Shown are the power spectra from 
the combined PN + MOS data (lower), the PN only (middle), and MOS only (top).
The best fitting power law + Lorentzian model is shown as the thick solid 
curve (lower).
\label{fig2}} 

\vskip 10pt

\figcaption[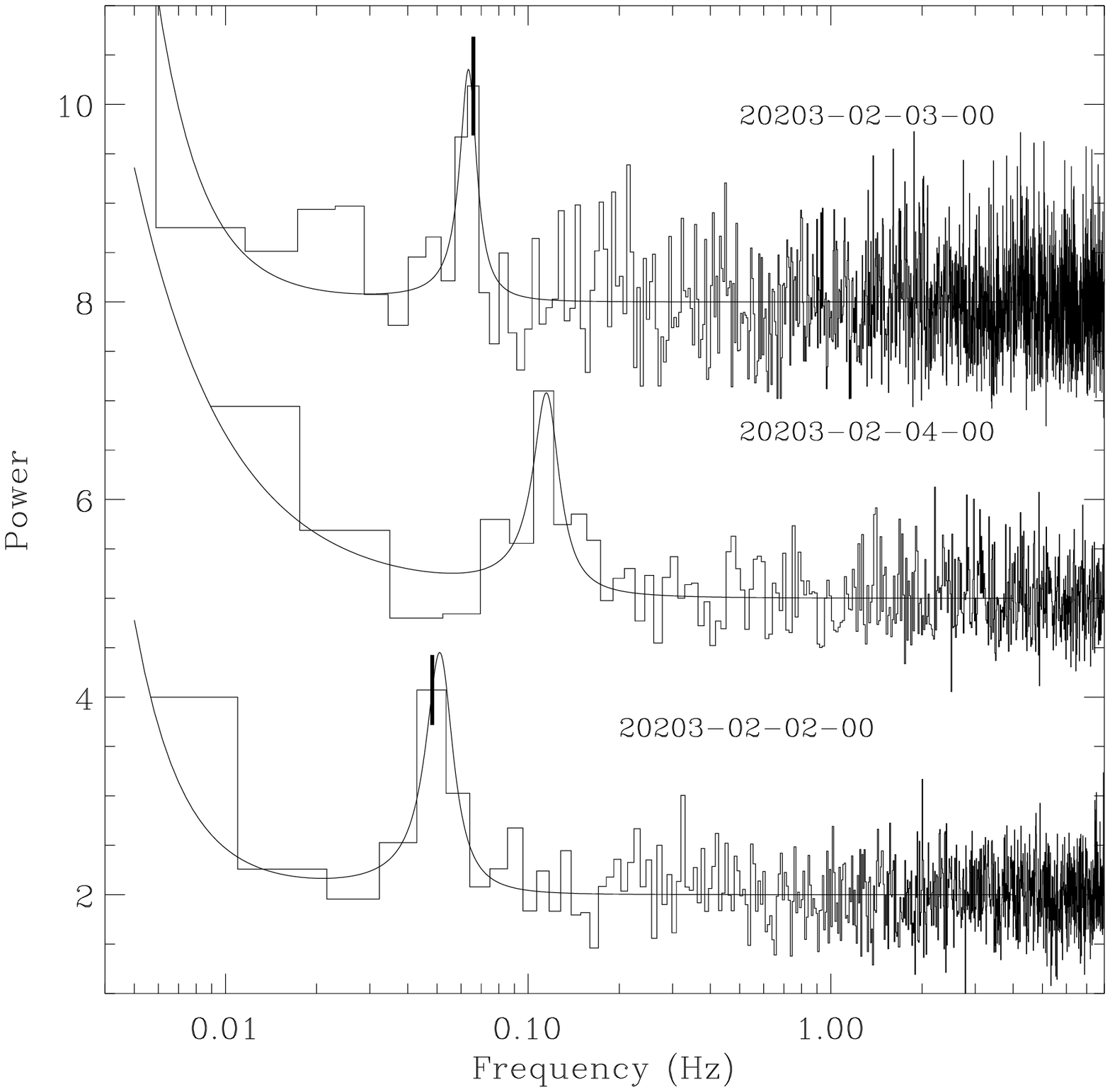]{Power spectra from three RXTE observations of M82 which 
show low frequency QPOs. Each spectrum is labelled with the corresponding 
OBSID. The best fit models are also shown.  For each spectrum a typical 
error bar near the peak of the QPO feature is shown.  The error bar for the 
middle curve (20203-02-04-00) has the same size as that shown for the bottom
curve (20203-02-02-00). 
\label{fig3}} 

\vskip 10pt

\figcaption[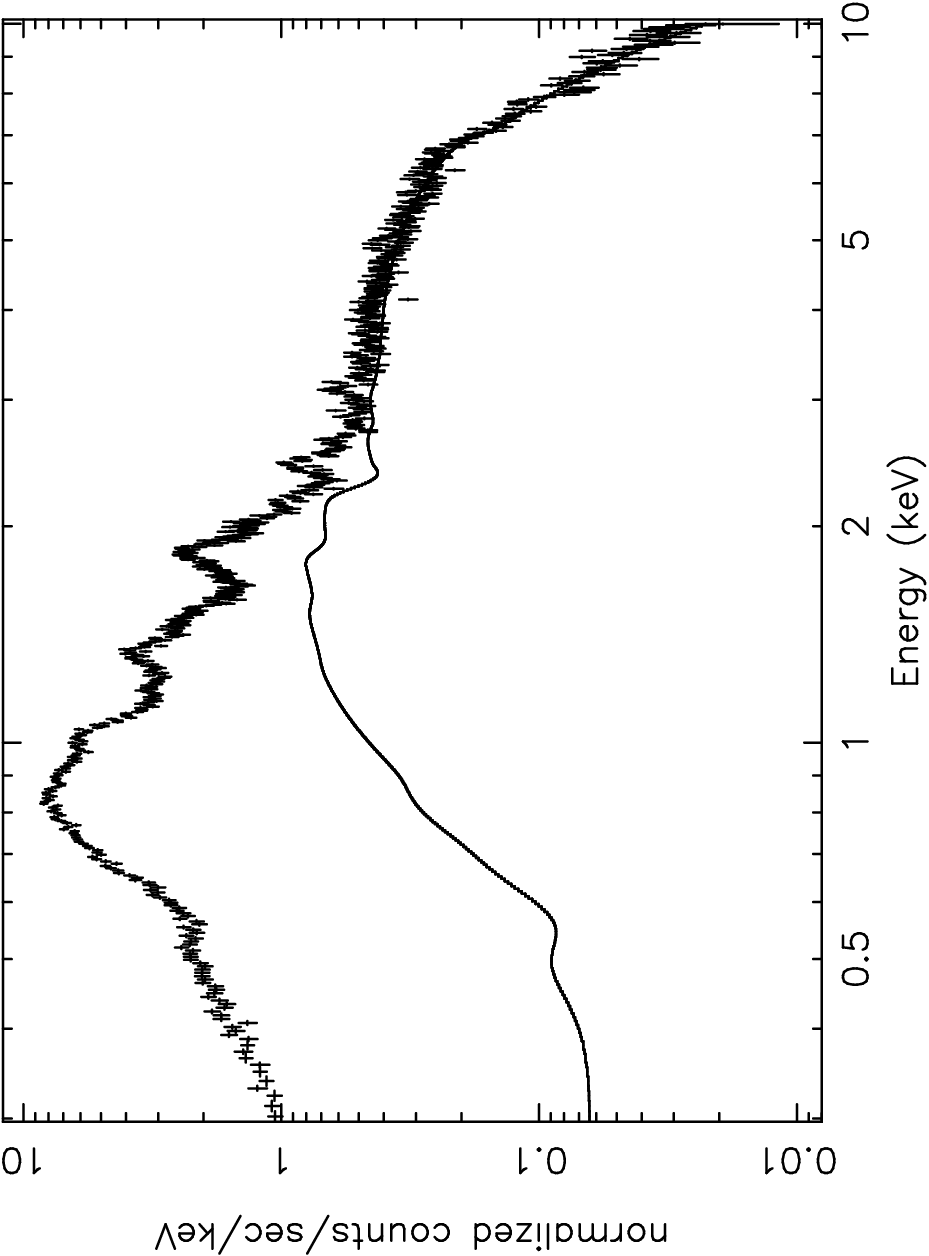]{The countrate spectrum in the PN inside the 18'' extraction
radius. The soft excess is from the diffuse thermal emission, and was
constrained based on Chandra observations which resolved the diffuse
component. The emission above 2 keV is dominated by the point source.
\label{fig4}} 

\vfill\eject

\newpage

\clearpage

\clearpage

\begin{figure}
\begin{center}
 \includegraphics[width=6in, height=6in]{f1.ps}
\end{center}
Figure 1: EPIC PN + MOS lightcurve of the ULX near the central region of M82. 
The time bins are 64 seconds.  A characteristic error bar is also shown.  
\end{figure}

\clearpage

\begin{figure}
\begin{center}
 \includegraphics[width=6in, height=6in]{f2.ps}
\end{center}
Figure 2: Power spectrum of the EPIC $> 2$ keV data from 
X41.4+60. The Nyquist frequency is 1 Hz. The frequency resolution is 4.7 mHz. 
The Poisson level has not been subtracted. Shown are the power spectra from 
the combined PN + MOS data (lower), the PN only (middle), and MOS only (top).
The best fitting power law + Lorentzian model is shown as the thick solid 
curve (lower).
\end{figure}

\clearpage

\begin{figure}
\begin{center}
 \includegraphics[width=6in, height=6in]{f3.ps}
\end{center}
Figure 3: Power spectra from three RXTE observations of M82 which 
show low frequency QPOs. Each spectrum is labelled with the corresponding 
OBSID. The best fit models are also shown.  For each spectrum a typical 
error bar near the peak of the QPO feature is shown.  The error bar for the 
middle curve (20203-02-04-00) has the same size as that shown for the bottom
curve (20203-02-02-00). 
\end{figure}

\begin{figure}
\begin{center}
 \includegraphics[width=6in, height=6in, angle=-90]{f4.ps}
\end{center}
Figure 4: The countrate spectrum in the PN inside the 18'' extraction
radius. The soft excess is from the diffuse thermal emission, and was
constrained based on Chandra observations which resolved the diffuse
component. The emission above 2 keV is dominated by the point source.
\end{figure}

\end{document}